\documentclass[english,PRL,two column,superscriptaddress,amsmath]{revtex4-1}
\usepackage[T1]{fontenc}
\usepackage[latin9]{inputenc}
\usepackage{babel}
\usepackage{amstext}
\usepackage{graphicx}
\usepackage[unicode=true,pdfusetitle,
 bookmarks=true,bookmarksnumbered=false,bookmarksopen=false,
 breaklinks=false,pdfborder={0 0 1},backref=false,colorlinks=false]
 {hyperref}
\hypersetup{
 colorlinks,linkcolor=red,citecolor=blue}
\usepackage{breakurl}

\makeatletter
\@ifundefined{textcolor}{}
{%
 \definecolor{BLACK}{gray}{0}
 \definecolor{WHITE}{gray}{1}
 \definecolor{RED}{rgb}{1,0,0}
 \definecolor{GREEN}{rgb}{0,1,0}
 \definecolor{BLUE}{rgb}{0,0,1}
 \definecolor{CYAN}{cmyk}{1,0,0,0}
 \definecolor{MAGENTA}{cmyk}{0,1,0,0}
 \definecolor{YELLOW}{cmyk}{0,0,1,0}
}

\makeatother

\begin{document}

\preprint{This line only printed with preprint option }

\title{Broadband frequency conversion via adiabatically tapered $\chi^{(2)}$ waveguide in photonic integrated circuits}

\author{Xiao Xiong}

\address{Key Laboratory of Quantum Information, University of Science and
Technology of China, CAS, Hefei, Anhui 230026, China}

\address{Synergetic Innovation Center of Quantum Information \& Quantum Physics,
University of Science and Technology of China, Hefei, Anhui 230026,
China}

\author{Chang-Ling Zou}

\address{Key Laboratory of Quantum Information, University of Science and
Technology of China, CAS, Hefei, Anhui 230026, China}

\address{Synergetic Innovation Center of Quantum Information \& Quantum Physics,
University of Science and Technology of China, Hefei, Anhui 230026,
China}

\address{Department of Applied Physics, Yale University, New Haven, Connecticut
06511, USA}

\address{Email: clzou321@ustc.edu.cn}

\author{Xiang Guo}

\address{Department of Applied Physics, Yale University, New Haven, Connecticut
06511, USA}

\author{Hong X. Tang}

\address{Department of Applied Physics, Yale University, New Haven, Connecticut
06511, USA}

\author{Xi-Feng Ren}

\thanks{Corresponding author: renxf@ustc.edu.cn}

\address{Key Laboratory of Quantum Information, University of Science and
Technology of China, CAS, Hefei, Anhui 230026, China}

\address{Synergetic Innovation Center of Quantum Information \& Quantum Physics,
University of Science and Technology of China, Hefei, Anhui 230026,
China}

\author{Guang-Can Guo}

\address{Key Laboratory of Quantum Information, University of Science and
Technology of China, CAS, Hefei, Anhui 230026, China}

\address{Synergetic Innovation Center of Quantum Information \& Quantum Physics,
University of Science and Technology of China, Hefei, Anhui 230026,
China}

\begin{abstract}
We propose to use the integrated aluminum nitride waveguide with engineered width variation to achieve optical frequency conversion based on $\chi^{(2)}$ nonlinear effect on a photonic chip. We show that in an adiabatically tapered waveguide, the frequency conversion has a much broader bandwidth and the efficiency within the bandwidth is almost constant, which is favorable for short pulses. We demonstrate both analytically and numerically an ``area law'' for the frequency conversion, i.e. the product of bandwidth and efficiency is conserved as long as peak conversion efficiency does not saturate. The adiabatic structure shows higher saturation thresholds in pump power or interaction length, outperforming the conventional uniform waveguide design. With our approach, high-efficiency and wavefront-keeping conversion for short pulses is possible on a photonic chip, which will surely find applications for scalable on-chip information processing.
\end{abstract}
\maketitle

\section{introduction}

Over last decades, there has been growing interests in the study of nonlinear optical processes in a photonic integrated circuit (PIC), in which the optical fields are tightly confined and the nonlinear effects are greatly enhanced \citep{wg,sphere,Sipe,Si,SiN-1,SiN-2,SiN-3,AlN-1,AlN-3}. Various applications based on $\chi^{(2)}$ or $\chi^{(3)}$ nonlinearities have been demonstrated, ranging from frequency conversion, frequency comb, super-continuum generation, parametric oscillation to quantum photon sources \citep{FC-2.4,comb1,comb2,comb3,corr1,corr2,indi,entangl1,entangl2}. Among these applications, coherent conversion between the signals at different frequencies is of great use, since the bandwidth of information processing would be extended by frequency multiplexing, thus allowing the hybridization of different systems \citep{quantum,coherent3,bridge,harald}. For example, the single photon conversion between visible and near-IR photons permits local high-fidelity quantum gate operation based on atoms, as well as high-efficiency communication between remote atoms.

However, the implementation of the nonlinear optical processes in PIC faces significant experimental challenges due to the requirement of phase-matching between different frequencies. In the PIC, the strong optical confinement also induces strong dispersion. Combining with the material dispersion effect, natural phase matching usually relies on precise structure design and fabrication. In practical, the phase-matching for a particular wavelength requires both coarse-tuning of the geometry by fabricating multiple devices with continuously varied sizes and fine-tuning of the dispersion by changing the temperature, which increases the experiment complexity. Additionally, the bandwidth of frequency conversion is also limited by the phase-matching condition, preventing the frequency conversion of ultrashort pulses \citep{pulse,SPM}.

In this Letter, we propose to convert the frequency of photons in an integrated waveguide with engineered waveguide geometry. Starting from nonlinear coupled-mode theory, we analyze the frequency conversion between $600\ \mathrm{nm}$ and $1550\ \mathrm{nm}$ with a pump laser at $980\ \mathrm{nm}$, bridging the visible and telecom wavelengths. Compared to the conventional phase-matched waveguide of uniform width, where conversion is only efficient in a very narrow frequency range, the conversion spectrum in adiabatically tapered waveguide is almost flat and offers much broader bandwidth \citep{adia-rev,rev-1,rev-2,adia2,adia3}. It is found that such broad bandwidth is obtained at the cost of peak efficiency, as the integration of conversion efficiency over the whole spectrum which we call it ``area'', maintains the same as that of phase matched process. However, when operated uniformly at high pump power, the adiabatic method outperforms the phase-matched method regarding both bandwidth and efficiency, because the conversion efficiency at center wavelength saturates for the phase-matched case. We further extend the adiabatic method to periodically modulated waveguide, the output spectrum of which features particular pattern. Such adiabatic waveguide geometry is more prone to fabrication imperfections, thus is very feasible for experimental implementation and can be used for frequency conversion for ultra-short pulses.

\section{Theory and model}

We start from the coupled-mode theory to investigate the frequency conversion process based on $\chi^{(2)}$-nonlinearity in integrated waveguide. The energy conservation relation between the photons involved is $\omega_{1}+\omega_{2}=\omega_{3}$, where subscripts $1,2,3$ stand for signal, pump and idler lights, respectively. For each frequency, the photons propagate in the waveguide as eigenmodes, whose wave-functions can be expressed as $\hat{E}_{m}(\vec{r},t)=u_{m}(x,y)e^{i\text{(}n_{m}k_{m}z-\omega_{m}t)}$ ($m=1,2,3$), with the effective modal index $n_{m}$ and the wave vector $k_{m}=\omega_{m}/c=2\pi/\lambda_{m}$. Here, $u_{m}(x,y)$ is the normalized transverse profile of the electric field, which satisfies $\int\int dxdy\hat{E}_{m}(x,y)\times\hat{H}^{*}_{m}(x,y)=\hbar\omega_{m}v_{g,m}$, where $\hat{H}_{m}$ corresponds to the magnetic field and $v_{g,m}$ is the group velocity \citep{norm}. Note here, we are using the total power flux for normalization rather than the total energy, so as to take into consideration the material dispersion \citep{flux,vg}. Then the operator of the total electric field is obtained as
\begin{equation}
\hat{\mathcal{E}}(\vec{r})=\sum_{m}\sqrt{\frac{\hbar\omega_{m}}{2}}(\hat{a}_{m}(z)u_{m}(x,y)e^{i\text{(}n_{m}k_{m}z-\omega_{m}t)}+h.c.),
\label{eq:E}
\end{equation}
where $\hat{a}_{m}(z)$ and their Hermitian conjugates are localized photon creation and annihilation operators.

Substituting Eq. (\ref{eq:E}) into Maxwell's equations and applying slowly varying amplitude approximation \citep{nonlinear-optics}, we obtain the coupled-mode equations for the $\chi^{(2)}$-process as
\begin{equation}
  \begin{split}
&\frac{\partial}{\partial z}\hat{a}_{1}=i\widetilde{g}\int\int
dxdyu_{1}^{*}u_{2}^{*}u_{3}\hat{a}_{2}^{\dagger}\hat{a}_{3}e^{-i\Delta\beta z},\\
&\frac{\partial}{\partial z}\hat{a}_{2}=i\widetilde{g}\int\int
dxdyu_{1}^{*}u_{2}^{*}u_{3}\hat{a}_{1}^{\dagger}\hat{a}_{3}e^{-i\Delta\beta z},\\
&\frac{\partial}{\partial z}\hat{a}_{3}=i\widetilde{g}\int\int
dxdyu_{1}u_{2}u_{3}^{*}\hat{a}_{1}\hat{a}_{2}e^{i\Delta\beta z}.
  \end{split}
\end{equation}
Here, $\widetilde{g}=2\pi\varepsilon_{0}\chi^{(2)}\sqrt{\frac{\hbar\omega_{1}\omega_{2}\omega_{3}}{2}}$, and $\Delta\beta=n_{1}k_{1}+n_{2}k_{2}-n_{3}k_{3}$ is the phase mismatch between signal, idler and pump modes. Under undepleted pump condition, the coupled-mode equations can be further reduced to be
\begin{equation}
  \begin{split}
&\frac{\partial}{\partial z}\hat{a}_{1}=ig^{*}\hat{a}_{3}e^{-i\Delta\beta z},\\
&\frac{\partial}{\partial z}\hat{a}_{3}=ig\hat{a}_{1}e^{i\Delta\beta z},\label{eq:cmt}
  \end{split}
\end{equation}
with the coupling strength $g=\left \langle \hat{a}_{2}\right \rangle\widetilde{g}\int\int dxdyu_{1}u_{2}u_{3}^{*}$. Here, $\left \langle \hat{a}_{2}\right \rangle=\sqrt{\frac{2\pi P_{2}}{\hbar\omega_{2}}}$ is the amplitude of classical pump field, and the integral corresponds to the field overlap between three modes. Note, though the pump light is treated classically to obtain analytic solution, the frequency conversion process is coherent, during which the quantum properties of photons are transduced \citep{coherent1,coherent2}. In the rotating coordinates ($\hat{A}_{1}=\hat{a}_{1}e^{i\frac{\Delta\beta}{2}z},\: \hat{A}_{3}=\hat{a}_{3}e^{-i\frac{\Delta\beta}{2}z}$),
the Hamiltonian of such coherent two-mode conversion in waveguide can be represented in matrix form as
\begin{equation}
\mathcal{H}(z)=\left(\begin{array}{cc}
\frac{\Delta\beta(z)}{2} & g^{*}(z)\\
g(z) & -\frac{\Delta\beta(z)}{2}
\end{array}\right).\label{eq:H}
\end{equation}

For a uniform waveguide with length $L$, $\frac{\partial}{\partial z}\mathcal{H}(z)=0$, thus the conversion efficiency will be
\begin{equation}
\eta=\left|g\right|^{2}L^{2}\mathrm{sinc}^{2}(\sqrt{\Delta\beta^{2}/4+\left|g\right|^{2}}L),\label{uniform}
\end{equation}
which shows that the nonlinear conversion efficiency is dependent on the phase mismatch $\Delta\beta$. Instead of fixing $\Delta\beta=0$, which corresponds to the strict phase matching, we will adiabatically tune $\Delta\beta(z)$ from negative to positive, which can be realized by engineering the waveguide geometry.

\begin{figure}
\includegraphics[width=8.5cm]{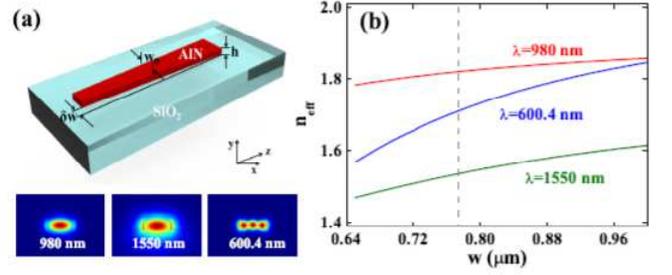}

\protect\caption{(Color online) (a) Schematic illustration of the adiabatic waveguide deposited on silica buffer, whose width varies
along light propagation direction as $w(z)$. Insets: the eigenmode distributions for signal, pump, and idler light, respectively,
with $w=760\ \mathrm{nm}$. (b) The effective refractive index $n_{eff}$ of eigenmodes at different wavelengths as a function of waveguide width $w$. Gray dashed line indicates the point where phase matching is satisfied.}
\label{fig1}
\end{figure}

Shown in Fig. 1(a) is the schematic of the tapered waveguide whose width $w$ is adiabatically varied along $z$-axis. The waveguide of height $h$ and length $L$ is on top of $\mathrm{SiO}_{2}$ buffer, with the waveguide width $w(z=\frac{L}{2})=w_{0}$ in the middle and width difference between input and output ports $\delta w$, respectively. So the width $w_{0}$ should satisfy $\Delta\beta(z=\frac{L}{2})=\Delta\beta(w_{0})=n_{1}(w_{0})k_{1}+n_{2}(w_{0})k_{2}-n_{3}(w_{0})k_{3}=0$ \citep{adia1}: as the waveguide width changes from $w<w_{0}$ to $w>w_{0}$, the phase mismatch is tuned from $\Delta\beta>0$ to $\Delta\beta<0$.

In this paper, we consider the aluminum nitride (AlN) as the waveguide material, which holds the advantages that is CMOS-compatible and possesses strong second-order nonlinearity ($\chi^{(2)}=4.7\ pm/V$). Additionally, AlN has enormous bandgap ($6.2\ \mathrm{eV}$) and allows operations ranging from ultraviolet (UV) up to infrared wavelengths. Therefore, AlN can be an excellent candidate to build CMOS-compatible PICs for quantum frequency conversion \citep{AlN-1,AlN-2,AlN-4,Xiang}. Then, the signal, pump and idler wavelengths are chosen to be $\lambda_{1}=1550\ \mathrm{nm}$, $\lambda_{2}=980\ \mathrm{nm}$, and $\lambda_{3}=600.4\ \mathrm{nm}$, respectively.

In the inset of Fig. 1(a), the electric field profiles of the modes involved in the frequency conversion process are also shown, where signal and pump light propagate as fundamental transverse-electric (TE) modes while idler light as the third-order TE mode. We should note that the first- and second-order modes for $\lambda_{3}=600.4\ \mathrm{nm}$ do exist as well, but they are not included in our model due to the large phase mismatch ($\Delta\beta\gg g$) and parity conservation. The modes are numerically calculated by finite element method (COMSOL Multiphysics) with practical parameters from experiments \citep{index}. To find $w_{0}$, we first studied the modal effective refractive indices as a function of waveguide width $w$ for the three eigenmodes, as displayed in Fig. 1(b). According to the gray dashed line which corresponds to $\Delta\beta(w)=0$, we obtain $w_{0}=0.773\ \mathrm{\mu m}$ for perfect phase matching. In the following studies, we'll carry out analytic calculations based on the effective refractive indices $n_{m}(w)$ fitted from Fig. 1(b).

Since both $\Delta\beta$ and $g$ vary with $w$, which is engineered along $z$-axis, the evolutions of $A_{m}$ will follow the Schr\"{o}dinger equation as
\begin{equation}
-i\frac{d}{dz}\left|\Phi(z)\right\rangle =\mathcal{H}(z)\left|\Phi(z)\right\rangle,
\label{schro}
\end{equation}
with $\left|\Phi(z)\right\rangle =\{A_{1}(z),\: A_{3}(z)\}^{T}$. Then the effect of $\mathcal{H}(z)$ on this system from $0$ to $z$
can be described with a transfer matrix $\mathcal{M}$, where
\begin{equation}
\left|\Phi(z)\right\rangle =\mathcal{M}\left|\Phi(0)\right\rangle =\mathcal{T} e^{i\int_{0}^{z}\mathcal{H}(s)ds}\left|\Phi(0)\right\rangle .
\end{equation}
Here, $\mathcal{T}$ is the position-ordering Dyson operator. Assuming the initial state is $\left|\Phi(0)\right\rangle =\{1,\:0\}^{T}$, $A_{3}$ in the final state becomes
$A_{3}(z)=\mathcal{M}_{21}$, where $\mathcal{M}_{21}$ is the off-diagonal matrix element of $\mathcal{M}$. Then the nonlinear
conversion efficiency is obtained as $\eta=\left|\frac{A_{3}(z)}{A_{1}(0)}\right|^{2}=\left|\mathcal{M}_{21}\right|^{2}$. Due to the conservation of photon numbers ($|A_{m}|^2$) during the frequency conversion, we have maximum efficiency $\eta=1$.
Here, the transfer matrix $\mathcal{M}=\mathcal{T}e^{i\int_{0}^{z}\mathcal{H}(s)ds}$ is numerically calculated with fourth-order Runge-Kutta method.

\section{Results}

\subsection{Parameter dependence}

\begin{figure}
\includegraphics[width=8.5cm]{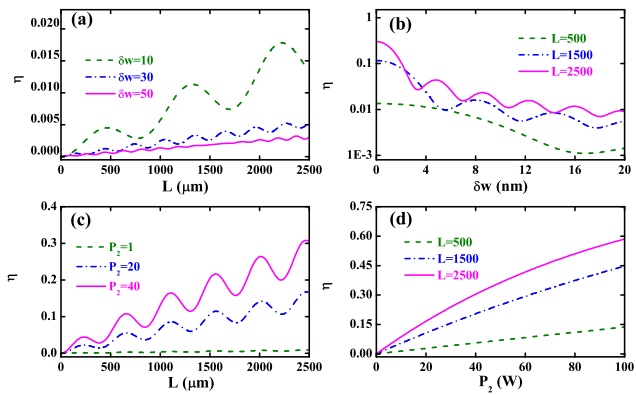}

\protect\caption{(Color online) (a) Dependence of $\eta$ on $L$, with $P_{2}=1\ \mathrm{W}$, $\delta w=10, 30, 50 \ \mathrm{nm}$,
respectively. (b) Dependence of $\eta$ on $\delta w$, with $P_{2}=1\ \mathrm{W}$, $L=500, 1500, 2500\ \mathrm{\mu
m}$, respectively. (c) Dependence of $\eta$ on $L$, with $\delta w=20\ \mathrm{nm}$, $P_{2}=1, 20, 40\ \mathrm{W}$,
respectively. (d) Dependence of $\eta$ on $P_{2}$, with $\delta w=20\ \mathrm{nm}$, $L=500, 1500, 2500
\ \mathrm{\mu m}$, respectively.}
\label{fig2}
\end{figure}

We first consider the waveguide width varies linearly along $z$-axis and is
expressed as $w(z)=w_{0}+\frac{\delta w}{L}(z-\frac{L}{2})$. Since the coupling strength $g$ depends on pump
power $P_{2}$, the nonlinear conversion efficiency $\eta$ is completely determined by $\delta w$, $L$, and $P_{2}$. Figure 2 illustrates
the dependence of $\eta$ on the three parameters. According to Fig. 2(a), $\eta$ increases with increasing $L$ overall, but
accompanied with oscillations. While the efficiency increases for small $\delta w$, the oscillation in the spectrum becomes stronger for small $\delta w$ as well, i.e. both the period and amplitude of the oscillation increase. When $\delta w$ reaches zero (phase matched case), $\eta$ will oscillate between 0 and 1 with a very long period. As we will
show later, the decrease of $\eta$ for large $\delta w$ is a compromise between efficiency and bandwidth. In Fig. 2(b),
the dependence of $\eta$ on $\delta w$ for different $L$ is displayed. When $\delta w=0$, which corresponds to the phase matched
case, the conversion efficiency is very high. Comparing the curves for different $L$, $\eta$ is higher for larger $L$, which is
consistent with the results in Fig. 2(a). When $\delta w>0$, $\eta$ falls off rapidly within $\delta w=5\ \mathrm{nm}$ (Note, the
coordinate of $\eta$ in Fig. 2(b) is Log scale). After that, it decreases smoothly with increasing $\delta w$. Figure 2(c) illustrates
$\eta$ as a function of $L$ with different pump power $P_{2}$. Again, $\eta$ grows as $L$ increases. It is also anticipated that the
efficiency is higher when the pump power is stronger. As for Fig. 2(d), it shows the dependence of $\eta$ on $P_{2}$ for different
waveguide length $L$. We can see that $\eta$ rises almost linearly with $P_{2}$ over certain ranges. According to these data, using a waveguide with larger $L$ and smaller $\delta w$, and pumping the nonlinear process with higher $P_{2}$ are preferred for the monochrome signal.

\subsection{Area law}

\begin{figure}
\includegraphics[width=8.5cm]{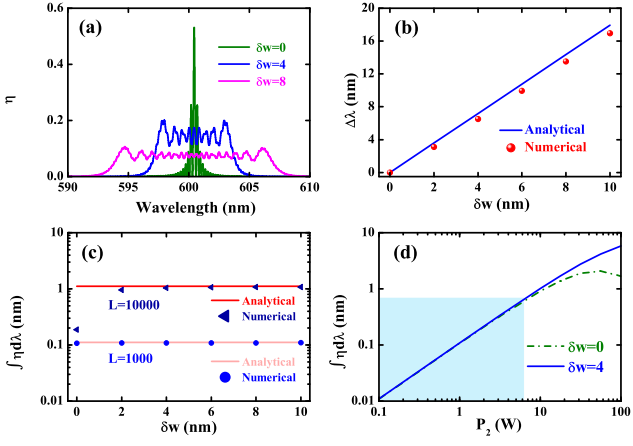}

\protect\caption{(Color online) (a) Nonlinear conversion spectra of $\lambda_{3}$ for $\delta w=0, 4, 8\ \mathrm{nm}$, with
$L=10000\ \mathrm{\mu m}$, $P_{2}=1\ \mathrm{W}$. (b) The bandwidth $\Delta\lambda$ extracted from (a), as the function of $\delta w$. (c)
Integration of $\eta$ over the whole spectrum as a function of $\delta w$ for $L=1000\ \mathrm{\mu m}$ and $L=10000\ \mathrm{\mu m}$, with $P_{2}=1\ \mathrm{W}$. (d) The integration as a function of $P_{2}$ for $\delta w=0$ and $\delta w=4\ \mathrm{nm}$, with $L=1000\ \mathrm{\mu m}$. The light blue region indicates where the ``area law'' holds ($P_{2}<6\ \mathrm{W}$).}
\label{fig3}
\end{figure}

To study the bandwidth of the frequency conversion in the adiabatically tapered waveguide, we calculated the spectra of $\lambda_{3}$ under different $\delta w$, as shown in Fig. 3(a). Apparently, when the pump wavelength is fixed, the center wavelength of spectra remains the same. The bandwidth of the adiabatic case can be engineered to be very broad. Therefore, it is quite suitable for frequency conversion of pulse laser, and the conversion between different states in frequency division multiplexing. Also, since the conversion efficiency $\eta$ within the bandwidth is almost flat, all frequency components of a pulse laser can be efficiently converted simultaneously. Thus, the waveform can be maintained. Note that, $P_{2}$ which is taken on the order of Watt, is actually reachable if a short pulsed pump is employed.

We extract the bandwidth $\Delta\lambda$ for different $\delta w$ from Fig. 3(a), and display it as red dots in Fig. 3(b). The blue line is obtained from analytic calculation. For an adiabatic waveguide, the difference of phase mismatch between input and output ports $\delta(\Delta\beta)$ is proportional to $\delta w$. Recalling the definition of phase mismatch $\Delta\beta=n_{1}k_{1}+n_{2}k_{2}-n_{3}k_{3}$, and differentiating the target wavelength ($\lambda_3$), we have
$\frac{d(\Delta\beta)}{d\lambda}=2\pi(n_3-n_1)/\lambda_3^2$.
Therefore, the bandwidth can be obtained from
\begin{equation}
\Delta\lambda\approx\delta w\cdot\frac{d(\Delta\beta)}{dw}/ \frac{d(\Delta\beta)}{d\lambda},
\end{equation}
which is proportional to $\delta w$.

As indicated by Fig. 3(a), there exists a trade-off between peak conversion efficiency and bandwidth. In the following, we show that this limit is only applicable to adiabatically tapered waveguide at low conversion efficiency regime (for example, in the case of weak pump or short interaction distance). We integrate $\eta$ over the whole
spectrum for different $\delta w$, with $L=1000\ \mathrm{\mu m}$ and $L=10000\ \mathrm{\mu m}$, respectively. As shown in Fig. 3(c), for $L=1000\ \mathrm{\mu m}$ (the blue dots), this integration is almost the same for uniform ($\delta w=0$) and adiabatic ($\delta w>0$) waveguides. This is what we call ``area law'', and it has proper analytic formula provided below.

For uniform waveguide with $\delta w=0$, according to Eq. (\ref{uniform}), the integration of $\eta$ over the whole spectrum results in
\begin{equation}
\int\eta d\lambda\approx2\pi\left|g\right|^{2}L/\frac{d(\Delta\beta)}{d\lambda},
\label{dw=0}
\end{equation}
under weak coupling $gL\ll1$. When $\delta w>0$, according to Landau-Zener (LZ) transition \citep{adia1}, the transition probability, i.e. the frequency conversion efficiency, can be expressed as \citep{adia-rev}
\begin{equation}
\eta=1-e^{-\frac{2\pi\left|g\right|^{2}}{\left|d\Delta\beta/dz\right|}}
\approx\frac{2\pi\left|g\right|^{2}}{\left|d\Delta\beta/dz\right|},
\label{lz}
\end{equation}
under the condition $\left|g\right|^{2} \ll \left|d\Delta\beta/dz\right|$. We can also obtain the integration asymptotically as
\begin{equation}
\int\eta d\lambda\approx\int\frac{2\pi\left|g\right|^{2}}{d(\Delta\beta)}d\lambda dz
=2\pi\left|g\right|^{2}L/\frac{d(\Delta\beta)}{d\lambda},
\label{dw>0}
\end{equation}
which is the same as Eq. (\ref{dw=0}). Eq. (\ref{dw>0}) is approximate because $g$ slightly changes
along $z$-axis for the adiabatic waveguide. The analytic $\int\eta d\lambda=0.1114$ for $L=1000\ \mathrm{\mu m}$, is plotted as light red line in Fig. 3(c), and demonstrates good agreement with the
numerical calculations.

This agreement may suggest that an adiabatic
waveguide may not be better than a uniform waveguide for wavelength conversions. However, this picture changes if we move on to the case $L=10000\ \mathrm{\mu m}$. Shown as triangles in Fig. 3(c), $\int\eta d\lambda$ for $\delta w>0$ is much larger than that for $\delta w=0$. The analytic $\int\eta d\lambda$ obtained from Eq. (\ref{dw>0}) is plotted as the red line in Fig. 3(c). The discrepancy
at $\delta w=0$ originates from the saturation of the peak conversion efficiency and can be interpreted more clearly combined with Fig. 3(d). In Fig. 3(d), the integrations for $\delta w=0$ and $\delta w=4\ \mathrm{nm}$ are plotted against pump power $P_{2}$. When the pump is weak, the peak $\eta$ are very small for both uniform and adiabatic waveguides.
The ``area law'' perfectly holds within the light blue region in Fig. 3(d): the integrations for $\delta w=0$ and $\delta w>0$ are equal, and are linearly dependent on $P_{2}$ since
$\left|g\right|^{2}\propto P_{2}$. As the pump power is stronger ($P_{2}>6\ \mathrm{W}$), the linear dependence breaks down for the uniform
waveguide. Because when $\delta w=0$, the peak $\eta$ at $\lambda_{3}=600.4 \mathrm{nm}$ which contributes most to the integration becomes saturated. On the contrary, for the adiabatic waveguide, the peak $\eta$ is smaller, resulting in a
higher saturation threshold $P_{2th}$. As $P_{2}$ increases (before saturation), the bandwidth remains the same. Thus, the
integration is still proportional to $P_{2}$. Similarly, when $L$ increases, the adiabatic waveguide also has a larger saturation
threshold $L_{th}$. That's why the integration for adiabatic waveguide is larger than that for uniform waveguide when $L=10000\ \mathrm{\mu m}$.

So far, we have proved that the adiabatically tapered waveguide outperforms the
uniform waveguide if the pump power is high enough or the waveguide is long enough, and the ``area law'' only holds at low conversion efficiency regime (no saturation). On the other hand, Figs. 3(c) and 3(d) also imply that the linear dependence of
$\int\eta d\lambda$ on $P_{2}$ and $L$ (Eq. (\ref{dw>0})) does not hold for adiabatic waveguide either beyond a threshold pump power ($P_{2th}^{'}$) or threshold length ($L_{th}^{'}$). At these new thresholds, $\left|g\right|^{2}$ (or $\left|d\Delta\beta/dz\right|$) becomes so large (or small) that the assumption $\left|g\right|^{2} \ll
\left|d\Delta\beta/dz\right|$ is not satisfied. We can no longer apply the LZ theory to simply estimate the efficiency as linear
function of $\left|g\right|^{2}$ and $L$ (Eq. (\ref{lz})), in which case the waveguide is more like a uniform waveguide.

\subsection{Periodically modulated waveguide}

\begin{figure}
\includegraphics[width=8.5cm]{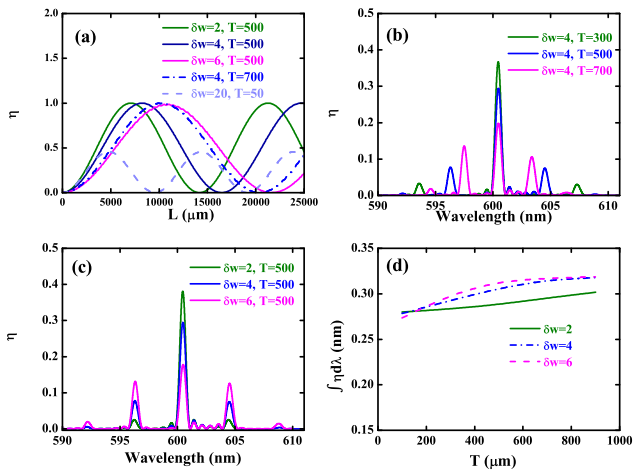}

\protect\caption{(Color online) (a) Conversion efficiency $\eta$ against $L$ for periodically modulated waveguide, with
$P_{2}=1\ \mathrm{W}$, $T=300, 500, 700\ \mathrm{\mu m}$, and $\delta w=2, 4, 6\ \mathrm{nm}$, respectively. (b) Nonlinear conversion spectra of $\lambda_{3}$ for periodically modulated waveguide, with
$P_{2}=1\ \mathrm{W}$, $\delta w=4\ \mathrm{nm}$, $L=3000\ \mathrm{\mu m}$, and $T=300, 500, 700\ \mathrm{\mu m}$,
respectively. (c) Nonlinear conversion spectra of $\lambda_{3}$ for periodically modulated waveguide, with
$P_{2}=1\ \mathrm{W}$, $T=500\ \mathrm{\mu m}$, $L=3000\ \mathrm{\mu m}$, and $\delta w=2, 4, 6\ \mathrm{nm}$,
respectively. (d) Integration of $\eta$ over the whole spectrum as a function of $T$, for periodically modulated waveguide with different $\delta w$.}

\label{fig4}
\end{figure}

Based on the results obtained above, we expect to improve the conversion efficiency with more flexible waveguide designs. In the following, we
will discuss the nonlinear frequency conversion processes in periodically modulated waveguide, whose width along $z$-axis is
$w(z)=w_{0}-\frac{\delta w}{2} cos(\frac{2\pi}{T}z)$. In Fig. 4(a), $\eta$ is shown as a function of interaction length $L$ with different waveguide width difference $\delta w$ and modulation period $T$. All these curves are oscillating but with a different period. For very small $\delta w$, $\eta$ can reach 1 (complete conversion). As $\delta w$ increases (the dashed light blue curve), the maximum achievable $\eta$ decreases, consistent with Fig. 2(b). Figure 4(a) implies that, when $L$ and $T$ are fixed, $\eta$ can be improved by adjusting $\delta w$. In the meantime, when $L$ and $\delta w$ are fixed, $\eta$ can also be improved by choosing an optimal $T$.

It's clear that with smaller $\delta w$ and smaller $T$, $\eta$ will reach 1 within shorter interaction length. However, this does not mean that $\delta w$ and $T$ should be made as small as possible. If we look into the spectrum for different $\delta w$ and $T$, the increase of $\eta$ will sacrifice the conversion efficiencies at sidebands, which is problematic for pulse operation. As shown in Fig. 4(b), when $\delta w$ is fixed as $4\ \mathrm{nm}$, the bandwidths of the spectrum envelopes are almost the same, but the sideband appears at different offset when $T$ is varied. This resonant characteristic in the spectrum can be explained as follows. During each period $T$, the nonlinear process is fully described with transfer matrix $\mathcal{M}$, which contains both phase and amplitude information. The nonlinear effect over the whole waveguide length $L$ is a coherent action of each $\mathcal{M}$, which is wavelength-dependent. Only when $\mathcal{M}$ fulfills constructive interference condition, the conversion efficiency $\eta$ will accumulate. Therefore, the peaks in the spectrum come out at specific wavelengths. In contrast, in Fig. 4(c), when $T$ is fixed, the peaks appear at the same wavelengths. Furthermore, when $\delta w$ gets smaller, the number of resonant wavelengths is also decreased. Combining Figs. 4(b) and 4(c), we find that $\delta w$ determines the bandwidth of the spectrum envelope, while the wavelengths that can fulfill constructive interference and get efficient conversion depend on $T$.

Finally, the conversion efficiency is integrated over the whole spectrum, as a function of $T$ for different $\delta w$ (Fig. 4(d)). Regardless of the minor difference at small $T$ and $\delta w$ where saturation may happen, the ``area law'' still applies for periodically modulated
waveguide, indicating the trade-off between peak conversion efficiency and operation bandwidth, which needs to be considered for specific situations, for example, monochromatic light or pulse operations.

It is notable that, the periodically modulated waveguide here is different from the quasi-phase matched (QPM) cases where waveguide width is also periodically modulated \citep{QPM1, QPM2, QPM3}. For QPM cases, the periodic modulation works like grating, whose period is designed to provide specific extra momentum kick and depends on the incident wavelength. While in our adiabatic case, the period is far insufficient to compensate the phase mismatch, and the conversion is also less sensitive to wavelength. Nevertheless, these two configurations are indeed relevant. When the modulation period is small enough to compensate the phase mismatch, the adiabatic case goes into the QPM regime. At large modulation period, the adiabatic case is analogous to QPM with chirped grating \citep{QPM1}.

\section{Discussion}

Finally, with all the results presented above, there are still several perspectives that we want to discuss and emphasize.

(1) The applicability of adiabatically tapered waveguide for frequency conversion. The principle of the waveguide geometry engineering is universal and can be applied to other nonlinear optical processes. For example, the second harmonic generation and spontaneous parametric down conversion. These coherent multiphoton interactions may find applications in scalable quantum computing system \citep{coherent3,bridge,QFC}. This design can be extended to other nonlinear waveguide materials, such as Si and SiN, enabling broadband four-wave mixing processes.

(2) The operation wavelengths. In this work, we choose the frequency conversion that bridges the visible and near-infrared wavebands, which has two-fold benefits. On one hand, telecom photons can be converted to the visible wavelength, enabling very high-efficiency detection with the commercial detector. On the other hand, we can also convert the emission from high-yield quantum dots, nitrogen-vacancy (NV) centers and silicon-vacancy (SiV) centers to telecom wavelength, to permit the long-distance quantum state transferring. One could also choose different pump wavelength to convert the photons to any desired wavelength.

(3) Experimental feasibility. The bandwidth of the adiabatic process is broader than that of the phase-matched case, and the almost constant efficiency over the whole bandwidth makes it very suitable for broadband ultrashort pulse operations. The broad output spectrum also means that we can get an efficient conversion at a particular wavelength with a broadband signal laser. Therefore, the requirements on incident wavelength and fabrication error due to phase matching condition are relaxed.

(4) Measuring the material nonlinear coefficient. Generally speaking, the nonlinear conversion efficiency strongly depends on the wavelength. So, it's hard to determine the nonlinear coefficient of a material according to the conversion efficiency at a particular wavelength, especially when the fabrication error presents. With adiabatically tapered waveguide, we can access to this coefficient according to the ``area law'', which is a very convenient feature of our model.

(5) Waveguide propagation loss. For a practical waveguide, the propagation loss is inevitable. The loss coefficient of a typical AlN waveguide is $\alpha \sim 1\ \mathrm{m^{-1}}$. According to our calculations, such loss almost does not dramatically reduce the nonlinear conversion efficiency within $L \sim 0.1\alpha^{-1}\ \mathrm{\mu m}$. The spectrum bandwidth of frequency conversion maintains the same too, with the efficiency slightly reduced in a whole.

\section{conclusion}

In summary, we apply the adiabatic method to on-chip $\chi^{(2)}$ nonlinear optical process in AlN waveguides. By controlling the waveguide width adiabatically, the frequency conversion has nearly constant efficiency within a broad bandwidth (hundred $\mathrm{nm}$ achievable). Even though the bandwidth broadening is at the cost of peak conversion efficiency, this configuration will benefit the conversion of pulses, where the peak power can be very high, but the broadband operation is desired. Moreover, we find that at low conversion efficiency regime (no saturation), the integration of efficiency over the whole bandwidth maintains the same as that for the perfectly phase-matched case, and linearly depends on the waveguide length and pump power. This ``area law'' nicely describes the trade-off between peak efficiency and bandwidth. However, due to the saturation of peak conversion efficiency, the linear dependence is easier to break down for the uniform waveguide. This means the adiabatic method is more favorable when the pump power is very strong. For example, we can use pulse to achieve both relatively high-frequency conversion efficiency and broad bandwidth.

Besides, we also extend the simple adiabatic waveguide to be periodically modulated structures, where the broad bandwidth and ``area law'' still hold. But its spectrum shows resonant feature due to the coherent interference between repeating segments, which provides another degree of freedom to control the output spectrum.

Our proposed adiabatic converter structure enables broad bandwidth, relaxes the phase matching condition, and is robust against the experimental imperfections, such as fabrication error and the variance of incident wavelength. Such compact structure will find applications for scalable on-chip information processing technology, especially for ultra-fast pulse optics.

\begin{acknowledgments}
This work was funded by the National Key R \& D Program (Grant No. 2016YFA0301700), Innovation Funds from the Chinese Academy of Sciences (Grant No. 60921091); NNSFC (Grant Nos. 11374289, 61590932, 61505195); the Fundamental Research Funds for the Central Universities, the Open Fund of the State Key Laboratory on Integrated Optoelectronics (IOSKL2015KF12). Hong X. Tang acknowledges supports from the David \& Lucile Packard Fellowship in Science and Engineering.
\end{acknowledgments}

\end{document}